\documentclass[preprint,aps]{revtex4}
\usepackage{graphicx}
\newcommand{\be}{\begin{equation}}
\newcommand{\ee}{\end{equation}}
\newcommand{\ba}{\begin{eqnarray}}
\newcommand{\ea}{\end{eqnarray}}

\newcommand{\f}{\frac}

\begin{document}
\title
{Formulation of general dynamical invariants and their unitary relations for time-dependent coupled quantum oscillators
 \vspace{0.3cm}}
\author{Jeong Ryeol Choi\footnote{E-mail: choiardor@hanmail.net } \vspace{0.3cm}}

\affiliation{Department of Nanoengineering, Kyonggi University, 
Yeongtong-gu, Suwon,
Gyeonggi-do 16227, Republic of Korea \vspace{0.7cm}}

\begin{abstract}
\indent
An exact invariant operator of time-dependent coupled oscillators is derived
using the Liouville-von Neumann equation.
The unitary relation between this invariant and the invariant
of two uncoupled simple harmonic oscillators is represented.
If we consider the fact that quantum solutions
of the simple harmonic oscillator is well-known,
this unitary relation is very useful in clarifying
quantum characteristics of the original
systems, such as entanglement, probability densities, fluctuations of the canonical variables,
and decoherence.
We can identify such quantum characteristics
by inversely transforming the mathematical representations
of quantum quantities belonging to the simple harmonic oscillators.
As a case in point, the eigenfunctions
of the invariant operator in the original systems are found through
inverse transformation of the well-known eigenfunctions associated with the simple harmonic oscillators.
\\
\\
{\bf Keywords}: coupled oscillators; invariant operator; unitary transformation; eigenfunction;
Hamiltonian
\end{abstract}

\maketitle
\newpage

{\ \ \ } \\
{\bf 1. Introduction
\vspace{0.2cm}}
\\
The interest in
dynamical and statistical properties of coupled oscillators has increased recently
mainly because they can be applied to analyze entanglement between next-generation
nano-optical systems \cite{aa1,aa2,aa3}.
The understanding and control of entanglement between nano-devices are crucial
in realizing diverse modern quantum technologies.
The dynamics of entanglement for coupled oscillators has been widely studied so far
from the fundamental quantum-mechanical point of view \cite{ac1,ac2,ac3,ac4,DP1}.
Besides nano-optics, other scientific disciplines,
such as electromagnetic induced transparency \cite{ab,ab2},
Josephson phenomena \cite{ef,ef2}, trapping of identical or different particles \cite{ME}, neural control of locomotion \cite{ncl},
and periodicity
of solar activity \cite{saa}, can also be analyzed by introducing a model of coupled oscillators.

If we regard complexity of the motion of coupled oscillators,
a rigorous treatment for it may be necessary especially when the time-dependence of
parameters are not so simple and/or the number of coupled oscillatory devices is more than two \cite{aa3}.
The introduction of the invariant operator theory is one of the methods that develop
quantum theory of time-dependent coupled oscillators in this regard.
The invariant operators can be formulated using the Lewis-Riesenfeld theory \cite{Lewis1,Lewis2}.
For non-coupled time-dependent oscillatory systems, the method of applying the invariant operator
in solving their quantum solutions is well known.
However, for the case of coupled time-varying oscillatory systems,
the decouplement between oscillators via an invariant may not be so easy
while it is necessary for deriving their complete quantum solutions.
Sometimes coupled oscillators with time-dependent parameters were treated by approximation
due to difficulty in associated mathematical developments \cite{rwa,rwa2,lah-3-7,ecr,ede}.
The exact formulation of the invariant is helpful in such a case.
We will establish a new formula of precise
invariant for two coupled time-dependent oscillators through this research.

The organization of this work is as follows.
In Sec. 2, we will formulate the
dynamical invariant based on its definition for coupled time-dependent oscillators,
starting from the Hamiltonian of the systems.
We will specify conditions in time variations of parameters, which are necessary for its formulation.
Unitary transformation method will be applied, in Sec. 3, in order to see how the obtained invariant operator is
related to the invariant operator of two simple harmonic oscillators (SHOs) that are mutually independent.
For this purpose, we first transform the invariant operator to a simple form by
means of appropriate unitary operators.
Then, we further transform the invariant operator by introducing a rotational unitary operator in order to diagonalize it.
The eigenfunctions of the invariant operator will be derived through these processes.
\\
\\
{\bf 2. Formulation of the Invariant \vspace{0.2cm}} \\
We consider the Hamiltonian of time-dependent coupled harmonic oscillators of the form
\be
\hat{H}(t) = \frac{1}{2}\sum_{j=1}^{2}\left[
\frac{\hat{p}_j^{2}}{m_j( t) }+b_j(t)(\hat{x}_j\hat{p}_j+\hat{p}_j\hat{x}_j)
+m_j( t) \omega _j^{2}(t) \hat{x}_{j}^{2}\right] + d(t)
\hat{x}_{1}\hat{x}_{2},   \label{1}
\ee%
where $m_j(t)$, $b_j(t)$, $\omega _j(t)$, and $d(t)$ are differentiable with respect to time.
This Hamiltonian has nonzero $b_j(t)$ and has not yet been treated, as far as we know,
if we consider the coupling together with the terms in Eq. (\ref{1}) and the
generality of time dependence of parameters
$m_j( t)$, $b_j(t)$, $\omega _j(t)$, and $d(t)$.
Some authors used the $b_j(t)$ term in order to analyze the behavior of interaction and damping
effects for Caldirola-Kanai oscillator coupled to two or three-level atoms
with the help of canonical transformations \cite{tla,tla2,tla3}.
This term also utilized in investigating squeezed states \cite{ssa},
in treating the master equation for damped oscillatory systems \cite{med},
in analyzing charged oscillator in a magnetic field \cite{com}, and so on.

From minor evaluations with the above Hamiltonian, we easily see that the classical
equations of motion of the systems are given by
\begin{eqnarray}
\ddot{x}_{1}+\f{\dot{m}_{1}}{m_{1}} \dot{x}_{1}+
\tilde{\omega}_{1}^{2}x_{1}
+ \f{d}{m_1} x_{2} &=&0,  \label{2} \\
\ddot{x}_{2}+\f{\dot{m}_{2}}{m_{2}} \dot{x}_{2}
+\tilde{\omega}_{2}^{2}x_{2}
+ \f{d}{m_2} x_{1} &=&0,  \label{3}
\end{eqnarray}%
where the modified angular frequencies are
\be
\tilde{\omega}_j^{2} = \omega_j^{2}-b_j^2-\dot{b}_j-b_j \f{\dot{m}_j}{m_j}.  \label{4}
\ee
Though the last term in Eq. (\ref{4}) is represented in terms of the time derivative of $m_j$,
the effects of the variations of masses may not be so large in most actual cases.

To formulate the invariant operator, we assume that it is of the form
\be
\hat{I} (t) = \f{1}{2}\sum_{j=1}^{2}\Big[
\alpha_j(t)  \hat{p}_j^{2}+\beta_j(t)  \left(
\hat{x}_j\hat{p}_j+\hat{p}_j\hat{x}_j\right)  +\gamma_j(t)  \hat{x}_j^{2}\Big]
+\delta(t) \hat{x}_{1}\hat{x}_{2}, \label{5}%
\ee
where $\alpha_j(t)$, $\beta_j(t)$, $\gamma_j(t)$, and $\delta(t)$ are coefficients that will be determined.
The dimension of $\hat{I} (t)$ can be any one.
For instance, the dimension of $\hat{I} (t)$ is energy for the case of Ref. \cite{cus},
square of angular momentum for the case of Ref. \cite{disn}, and velocity$\times$length for the case of Ref. \cite{cms}.
We take the dimension of $\hat{I}(t)$ as energy in our case for convenience.
Then, the dimension of $\alpha_j$, $\beta_j$, $\gamma_j$, and $\delta$ are
${\mathrm M}^{-1}$, ${\mathrm T}^{-1}$, ${\mathrm M}{\mathrm T}^{-2}$, and ${\mathrm M}{\mathrm T}^{-2}$, respectively.

We can derive the coefficients of the invariant operator using the Liouville-von Neumann equation
which designates that the time derivative of $\hat{I}$ should be zero:
\begin{equation}
\frac{d\hat{I}}{dt}=\frac{\partial \hat{I}}{\partial t}+ \f{1}{i\hbar}
[\hat{I}, \hat{H}]=0.  \label{6}
\end{equation}%
From the substitution of Eqs. (\ref{1}) and (\ref{5}) into this equation,
we easily have
\begin{equation}
\dot{\alpha}_j(t) =2b_j(t)\alpha_j(t)-\frac{2\beta_j( t) }{m_j(t)},
\label{7}
\end{equation}%
\begin{equation}
\dot{\beta}_j( t) =m_j( t)\alpha_j( t) \omega_j^{2}(t)  -\frac{\gamma_j(t)}{m_j(t)},
\label{8}
\end{equation}%
\begin{equation}
\dot{\gamma}_j(t)=-2b_j(t)\gamma_j(t)+2m_j( t) \beta_j(t) \omega _j^{2}(t) ,  \label{9}
\end{equation}%
\be
\dot{\delta}(t) =-\delta(t)[ b_{1}(t) +b_{2}(t) ]+d(t)[ \beta_{1}(t) +\beta_{2}(t) ] ,  \label{10} 
\ee%
\be
\f{\delta(t)}{d(t)} = F(t),  \label{11}
\ee
where $F(t)$ is given by
\be
F(t) = \alpha_1(t)m_1(t),  \label{12}
\ee
under the requirement
\be
\alpha_1(t)m_1(t) = \alpha_2(t)m_2(t).  \label{13}
\ee
The condition in Eq. (\ref{13}) does not mean $m_1(t)=m_2(t)$, because we can take
$\alpha_1(t) \neq \alpha_2(t)$.
On account of this, the invariant operator method is applicable without approximation
even when the two masses are different from each other.

To solve Eqs. (\ref{7})-(\ref{11}), we put $\alpha_j(t)$ as
\begin{equation}
\alpha_j(t)  =\alpha_{0,j}\rho_j^{2}(t), \label{14}%
\end{equation}
where $\alpha_{0,j}$ are arbitrary real constants and $\rho_j(t)$ are the solutions
of the following differential equations \cite{ssf,ssf2}
\begin{equation}
\ddot{\rho}_j+\frac{\dot{m}_j}{m_j}\dot{\rho}_j+\tilde{\omega}_j^2(t)  \rho_j
=\frac{\Omega_j^2}{4m_j^{2}\rho_j^{3}},  \label{15}
\end{equation}
whereas $\Omega_j$ are real constants with dimension of ${\mathrm M}{\mathrm L}^{2}{\mathrm T}^{-1}$.
Some authors putted $\Omega_j$ as 1
or a dimensionless number \cite{Lewis1,ese,ese2}.
Now it is possible to obtain $\beta_j(t)$ and $\gamma_j(t)$ from Eq. (\ref{7}) and Eq. (\ref{8}), respectively, as
\begin{equation}
\beta_j(t)=\alpha_{0,j}m_j(t)[b_j(t)\rho_j^2(t)- \rho_j(t)\dot{\rho}_j(t)],  \label{16}
\end{equation}%
\begin{equation}
\gamma_j(t)  =\alpha_{0,j} \Bigg[\f{\Omega_j^2}{4\rho_j^2(t)}
+ m_j^2(t) \Big(b_j^2(t)\rho_j^2(t)-2b_j(t)\rho_j(t)\dot{\rho}_j(t)+\dot{\rho}_j^2(t)\Big)\Bigg].  \label{17}
\end{equation}
We easily confirm that $\beta_j(t)$ and $\gamma_j(t)$ obtained in such a way satisfy Eq. (\ref{9}).
Using Eq. (\ref{11}) together with Eq. (\ref{10}), it is possible to represent $\delta(t)$ in the form
\be
\delta(t)=F(t)d(t),  \label{18}
\ee
under the requirement that $d(t)$ should follow the relation
\be
\dot{d}(t) = -G(t) d(t),  \label{19}
\ee
where
\be
G(t)= \f{\dot{m}_1(t)}{m_1(t)}+\f{3\dot{\rho}_1(t)}{\rho_1(t)}+\f{\dot{\rho}_2(t)}{\rho_2(t)}.  \label{20}
\ee
Some comments on deriving $G(t)$ are represented in Appendix A.
Because $G(t)$ is not expressed in terms of $b_j(t)$, the condition given in Eq. (\ref{19})
is necessary even when $b_j(t)=0$.

We now confirm that Eq. (\ref{5}) with Eqs. (\ref{14}), (\ref{16}), (\ref{17}), and (\ref{18}) is the invariant operator.
This invariant operator is valid under the two groups of conditions,
where the first group is given by Eq. (\ref{13}) and the second group is given by Eq. (19) with Eq. (\ref{20}).
The merit of our formula of $\hat{I}(t)$
is that it holds without any approximation provided the aforementioned two groups of conditions.
This exact dynamical invariant is useful in analyzing quantum
mechanical properties of the systems that we have considered.
In particular, the dynamical invariant can be used in analysis of the entanglement for
coupled oscillators \cite{DP1}.
\\
\\
{\bf 3. Unitary Relations \vspace{0.2cm}} \\
Not only the Hamiltonian but also the invariant operator described in the previous section
involves a coupling term (the last term in Eq. (\ref{5})).
Unitary transformation method is useful in such a case since it affords simplification of the
invariant operator through transformation \cite{ac1,lah-3-7,ede,cus,ssf2}.
Now we will decouple the coupling term in $\hat{I}(t)$ by successive several unitary transformations of it.
To this end, we first introduce the unitary operator of the form
\begin{equation}
\hat{U}_A=\hat{U}_{A1}\hat{U}_{A2},  \label{22}
\end{equation}%
where
\ba
\hat{U}_{A1}&=&\prod_{j=1}^{2}\exp \left( \frac{i}{2\hbar }(\hat{p}_j\hat{x}_j
+\hat{x}_j\hat{p}_j)\ln \sqrt{\f{1}{M\alpha_j(t)}}\right) ,
\label{23} \\
\hat{U}_{A2}&=&\exp \bigg( -\frac{i}{2\hbar }\sum_{j=1}^{2} M\beta_j( t) \hat{x}_j^{2}\bigg) ,
\label{24}
\ea%
whereas $M$ is a real constant with the dimension of ${\mathrm M}$.
The transformation of the original invariant operator can be performed by using the relation
\be
\hat{I}_{A}(t) = \hat{U}_{A}^{-1} \hat{I}(t) \hat{U}_{A}.  \label{21}
\ee
A little evaluation after inserting Eqs. (\ref{5}) and (\ref{22}) into the above equation gives the
formula of the transformed invariant operator, such that
\ba
\hat{I}_{A}( t) &=&\f{1}{2}\sum_{j=1}^{2}\bigg(
\f{\hat{p}_j^{2}}{M} + M \omega_{0,j}^2 \hat{x}_j^{2}\bigg) +M\delta(t)\sqrt{\alpha_1(t)\alpha_2(t)}
\hat{x}_{1}\hat{x}_{2},  \label{25}%
\ea
where
\be
\omega_{0,j}^2 = \alpha_j(t)\gamma_j(t)-\beta_j^2(t)=\f{\alpha_{0,j}^2\Omega_j^2}{4}.  \label{26}
\ee
If we consider that the term that involves a parenthesis in Eq. (\ref{25}) is the same as the Hamiltonian
of the SHOs of mass $M$ and angular frequencies $\omega_{0,j}$,
$\hat{I}_{A}( t)$ is simper than the original invariant operator.

We can further simplify the invariant operator by removing the last term in Eq. (\ref{25})
through a rotational transformation.
Taking notice of this, we consider
\be
\hat{I}_{B}(t) = \hat{U}_{B}^{-1} \hat{I}_{A}(t) \hat{U}_{B},  \label{27}
\ee
as the next transformation, where the unitary operator is of the form
\ba
\hat{U}_{B}&=&\exp\bigg( -\frac{i\varphi}{\hbar}\left( \hat{p}_{2}\hat{x}_{1}%
-\hat{p}_{1}\hat{x}_{2}\right)  \bigg). \label{28}%
\ea
A minor evaluation at this stage gives
\ba
\hat{I}_{B}( t) &=&\f{1}{2}\sum_{j=1}^{2}\bigg(
\f{\hat{p}_j^{2}}{M} + M \bar{\omega}_{0,j}^2 \hat{x}_j^{2}\bigg)
+\bar{\delta} \hat{x}_{1}\hat{x}_{2},  \label{29}%
\ea
where
\ba
\bar{\omega}_{0,1}^2 &=& \omega_{0,1}^2\cos^2\varphi+\omega_{0,2}^2\sin^2\varphi
+\delta(t)\sqrt{\alpha_1(t)\alpha_2(t)}\sin(2\varphi),  \label{30} \\
\bar{\omega}_{0,2}^2 &=& \omega_{0,1}^2\sin^2\varphi+\omega_{0,2}^2\cos^2\varphi
-\delta(t)\sqrt{\alpha_1(t)\alpha_2(t)}\sin(2\varphi),  \label{31} \\
\bar{\delta} &=& M\bigg(\delta(t)\sqrt{\alpha_1(t)\alpha_2(t)}\cos(2\varphi)
- \f{1}{2}(\omega_{0,1}^2-\omega_{0,2}^2)\sin(2\varphi) \bigg).  \label{32}
\ea
Now we take
\be
\varphi =\f{1}{2}\mathrm{atan} \left((\omega_{0,1}^2-\omega_{0,2}^2)/2, \delta(t)\sqrt{\alpha_1(t)\alpha_2(t)}\right),
  \label{33}
\ee
where $\vartheta \equiv \mathrm{atan}(z_1,z_2)$ is the two-variable arctangent function of $\tan\vartheta = z_2/z_1$.
Then the cross term which is the last term in Eq. (\ref{29}) disappears, leading
the finally transformed invariant operator in the form
\be
\hat{I}_{B} =\f{1}{2}\sum_{j=1}^{2}\bigg(
\f{\hat{p}_j^{2}}{M} + M \bar{\omega}_{0,j}^2 \hat{x}_j^{2}\bigg).  \label{34}%
\ee
Equations (\ref{30}), (\ref{31}), and (\ref{32}) are expressed in terms
of $\delta(t)\sqrt{\alpha_1(t)\alpha_2(t)}$ as one can see.
From direct differentiation of this function with respect to $t$ using
Eqs. (\ref{14}), (\ref{15}), (\ref{18}), (\ref{19}), and (\ref{20}),
we have $d\Big[\delta(t)\sqrt{\alpha_1(t)\alpha_2(t)}\Big]/dt =0$.
By combining this fact with the fact that $\omega_{0,j}$ are constants (see Eq. (\ref{26})),
we confirm that $\bar{\omega}_{0,j}$ and $\varphi$ (Eq. (\ref{33})) are constants.
Hence the eventual 
invariant $\hat{I}_{B}( t)$ is the same as the invariant operator of
two SHOs that are mutually independent.
Notice that, for the case of a SHO, the Hamiltonian itself is a quadratic invariant operator.

Based on basic quantum mechanics, let us now express Eq. (\ref{34}) in a way that
\be
\hat{I}_{B} =\sum_{j=1}^{2}\hbar \bar{\omega}_{0,j} \bigg(\hat{a}_{0,j}^\dagger \hat{a}_{0,j} + \f{1}{2}\bigg),
  \label{35}
\ee
where $\hat{a}_{0,j}$ are annihilation operators of the form
\be
\hat{a}_{0,j} = \sqrt{\f{M \bar{\omega}_{0,j}}{2\hbar}} \hat{x}_j + \f{i}{\sqrt{2M\bar{\omega}_{0,j}\hbar}}\hat{p}_j,
  \label{36}
\ee
whereas their hermitian adjoints, $\hat{a}_{0,j}^\dagger$, are the creation operators.
By the inverse transformation of Eq. (\ref{35}), we see that the original invariant operator given in Eq. (\ref{5})
can also be represented as
\be
\hat{I}(t) =\sum_{j=1}^{2}\hbar \bar{\omega}_{0,j} \bigg(\hat{a}_j^\dagger \hat{a}_j + \f{1}{2}\bigg),
  \label{37}
\ee
where $\hat{a}_j$ are annihilation operators in the original systems, which are related to $\hat{a}_{0,j}$ by
\be
\hat{a}_j = \hat{U}_{A} \hat{U}_{B}\hat{a}_{0,j}\hat{U}_{B}^{-1}\hat{U}_{A}^{-1},  \label{38}
\ee
and the hermitian adjoints $\hat{a}_j^\dagger$ are the corresponding creation operators.
The mathematical formulae of $\hat{a}_j$ are represented in Appendix B.

Let us write the eigenvalue equations for $\hat{I}_{B}$ as
\be
\hat{I}_{B} u_{0,n_{1},n_{2}}(x_1,x_2) = \lambda_{n_{1},n_{2}} u_{0,n_{1},n_{2}}(x_1,x_2),  \label{39}
\ee
where $\lambda_{n_{1},n_{2}}$ are eigenvalues and $u_{0,n_{1},n_{2}}(x_1,x_2)$ are eigenfunctions.
Then, we can readily express the solutions of Eq. (\ref{39}) in the form
\begin{equation}
u_{0,n_{1},n_{2}}(x_1,x_2) = \prod_{j=1}^{2}
\sqrt[4]{\frac{M\bar{\omega}_{0,j}}{\pi\hbar }}
\f{1}{\sqrt{2^{n_j}n_j!}}
H_{n_j}\left(  \sqrt{\frac{M\bar{\omega}_{0,j}}{\hbar}%
}x_j\right)  \exp\left[  -\frac{M\bar{\omega}_{0,j}}{2\hbar}x_j^{2}\right]  ,  \label{40}
\end{equation}
\be
\lambda_{n_{1},n_{2}} =\sum_{i=1}^{2}\hbar \bar{\omega}_{0,j} \bigg(n_j + \f{1}{2}\bigg).  \label{41}
\ee
If we write the eigenvalue equations for the original invariant operator as
\be
\hat{I}(t) u_{n_{1},n_{2}}(x_1,x_2,t) = \lambda_{n_{1},n_{2}} u_{n_{1},n_{2}}(x_1,x_2,t),  \label{42}
\ee
the corresponding eigenfunctions $u_{n_{1},n_{2}}(x_1,x_2,t)$ can be obtained from the
unitary relation:
\be
u_{n_{1},n_{2}}(x_1,x_2,t) = \hat{U} u_{0, n_{1},n_{2}}(x_1,x_2),  \label{43}
\ee
where
\be
\hat{U} = \hat{U}_A \hat{U}_B . \label{44}
\ee
The straightforward evaluation of Eq. (\ref{43}) leads to
\begin{equation}
u_{n_{1},n_{2}}(x_1,x_2,t) = \prod_{j=1}^{2}
\sqrt[4]{\frac{\bar{\omega}_{0,j}}{\pi\hbar\alpha_j(t) }}
\f{1}{\sqrt{2^{n_j}n_j!}}
H_{n_j}\left(  \sqrt{\frac{\bar{\omega}_{0,j}}{\hbar}%
}X_j\right)  \exp\left[  -\frac{1}{2\hbar}\bigg(\bar{\omega}_{0,j}X_j^{2}+i \f{\beta_j(t)}{\alpha_j(t)}x_j^{2} \bigg)\right]  ,  \label{45}
\end{equation}
where
\begin{equation}
\left(
\begin{array}
[c]{c}%
X_{1}\\
X_{2}%
\end{array}
\right)  =\left(
\begin{array}
[c]{cc}%
\frac{1}{\sqrt{\alpha_{1}(t)  }}\cos\varphi  &
\frac{1}{\sqrt{\alpha_{2}(t)  }}\sin\varphi
\\
-\frac{1}{\sqrt{\alpha_{1}(t)  }}\sin\varphi  &
\frac{1}{\sqrt{\alpha_{2} (t)  }}\cos\varphi
\end{array}
\right)  \left(
\begin{array}
[c]{c}%
x_{1}\\
x_{2}%
\end{array}
\right)  . \label{46}%
\end{equation}
Thus, we have obtained the full eigenfunctions of the original invariant operator.
These functions are Fock states, which are associated with a well-defined number of quanta
(or photons in quantum optics).
Because the solutions, Eq. (\ref{45}) together with Eq. (\ref{41}),
in the original systems do not represented in terms of $M$, the scale of $M$ does not affects
the quantum solutions as expected.

The unitary relations found here are useful in estimating quantum characteristics of
original systems.
Inverse transformations of quantum quantities belonging to the SHOs enable us to know those in original systems.
Based on this, we can find entanglement, fluctuations of canonical variables, Wigner distribution functions,
Mandel's Q parameter, decoherence property, etc.
\\
\\
{\bf 4. Conclusion
\vspace{0.2cm}}
\\
The quantum invariant operator for the two coupled time-dependent oscillators was evaluated.
The Hamiltonian of the systems that we have considered depends on time in the most arbitrary manner
so far as the restrictions imposed in the formula of the invariant operator allow.
We also investigated the unitary relation
between the original invariant operator and the invariant operator of two independent SHOs.
The obtained invariant operator is helpful in analyzing dynamical properties of
time-varying oscillatory systems.

The firstly transformed invariant operator given in Eq. (\ref{25}) is
simpler than the original invariant operator.
However, it still involves a coupling term $\hat{x}_1\hat{x}_2$.
Such a term has been finally removed by the second unitary transformation,
leading the invariant being no longer expressed in terms of time.
The transformed invariant operator is also represented by ladder operators of the SHOs.
By inversely transforming the eigenfunctions in the transformed systems to the ones
associated with the original systems,
we have identified the eigenfunctions of the original invariant operator.

The main contribution of this investigation is the formulation of the exact invariant operator
and demonstrating how to transform it to that of the SHOs.
Our research can be utilized in analysis of quantum properties of various dynamical systems
that are described by coupled oscillators, especially nano-optomechanical systems \cite{aa1,aa2,ede}.
Entanglement dynamics between two or multi-coupled nano-optomechnical oscillators
can be elucidated using invariants.
Additionally, vibrations of diatomic molecules \cite{cdm,cdm2},
array of electromechanical devices \cite{aed}, coupled autonomous dynamical systems \cite{cad},
and so on can also be analyzed utilizing our research.
Clarification of novel nonlocal effects in coupled quantum information systems starts
from the analysis of their fundamental quantum characteristics.

A recent trend in optomechanical and electromechanical systems is that the size of devices
gradually become smaller according to the advance of nano-based technologies.
Due to the prominence of quantum effects as the devices become small
towards nanoscale, the understanding of related quantum phenomena is important \cite{nq1}.
Our research may provide actual solutions for feasible quantum description of
coupled oscillatory nanosize systems via a dynamical invariant operator.

\appendix
\section{\bf About the formula of $G(t)$}
Equation (\ref{19}) which involves $G(t)$ is evaluated using Eq. (\ref{10}) together with (\ref{11}).
In this evaluation, the formula of $G(t)$ given in Eq. (\ref{20})
is derived using $F(t)=\alpha_1(t)m_1(t)$.
On the other hand, if we use $F(t)=\alpha_2(t)m_2(t)$ instead of it, we have
\be
G(t)= \f{\dot{m}_2(t)}{m_2(t)}+\f{\dot{\rho}_1(t)}{\rho_1(t)}+\f{3\dot{\rho}_2(t)}{\rho_2(t)}.   \label{A1}
\ee
Both the expressions of $G(t)$ in Eqs. (\ref{20}) and (\ref{A1}) are asymmetrical about $\rho_j(t)$ as we
can see.
However, we can also obtain a symmetrical expression of it by combining Eq. (\ref{20}) and Eq. (\ref{A1}) together.
The result of this procedure is
\be
G(t)= \f{1}{2} \Bigg(\f{\dot{m}_1(t)}{m_1(t)}+ \f{\dot{m}_2(t)}{m_2(t)}\Bigg)
+2\Bigg( \f{\dot{\rho}_1(t)}{\rho_1(t)}+\f{\dot{\rho}_2(t)}{\rho_2(t)} \Bigg).   \label{A2}
\ee

\section{\bf Full representation of $\hat{a}_j$}
It is possible to evaluate Eq. (\ref{38}) straightforwardly through the use of Eqs. (\ref{22}) and (\ref{28}).
This evaluation gives
\ba
\hat{a}_1 &=& \f{1}{\sqrt{2\hbar \bar{\omega}_{0,1}}} \bigg[ \bar{\omega}_{0,1} \bigg(\f{1}{\sqrt{\alpha_1}}\cos\varphi \hat{x}_1
+ \f{1}{\sqrt{\alpha_2}}\sin\varphi \hat{x}_2 \bigg) \nonumber \\
& &+i (\sqrt{\alpha_1}\cos\varphi \hat{{\textsf p}}_1
+\sqrt{\alpha_2}\sin\varphi \hat{{\textsf p}}_2)\bigg], \\
\hat{a}_2 &=& \f{1}{\sqrt{2\hbar \bar{\omega}_{0,2}}} \bigg[ \bar{\omega}_{0,2} \bigg(-\f{1}{\sqrt{\alpha_1}}\sin\varphi \hat{x}_1
+ \f{1}{\sqrt{\alpha_2}}\cos\varphi \hat{x}_2 \bigg) \nonumber \\
& &+i (-\sqrt{\alpha_1}\sin\varphi \hat{{\textsf p}}_1
+\sqrt{\alpha_2}\cos\varphi \hat{{\textsf p}}_2)\bigg],
\ea
where
\be
\hat{{\textsf p}}_j = \hat{p}_j + \f{\beta_j}{\alpha_j} \hat{x}_j.
\ee
\\
\\


\end{document}